\documentclass{article}

\usepackage{PRIMEarxiv}
\usepackage[T1]{fontenc}    
\usepackage{url}            
\usepackage{booktabs}       
\usepackage{amsfonts}       
\usepackage{nicefrac}       
\usepackage{microtype}      
\usepackage{lipsum}
\usepackage{fancyhdr}       
\usepackage{graphicx}       

\usepackage[version=3]{mhchem}
\usepackage{subcaption}
\usepackage{xcolor}
\usepackage{bm}
\usepackage{hyperref}
\usepackage{amssymb}
\usepackage[symbol]{footmisc}

\title{MLPROP -- an open interactive web interface for thermophysical property prediction \\with machine learning}

\author{
  Marco Hoffmann, Thomas Specht, Nicolas Hayer, Hans Hasse, Fabian Jirasek \\
  Laboratory of Engineering Thermodynamics (LTD) \\
  RPTU Kaiserslautern \\
  \texttt{fabian.jirasek@rptu.de} \\
}

\begin{document}
\maketitle

\begin{abstract}
Machine learning (ML) enables the development of powerful methods for predicting thermophysical properties with unprecedented scope and accuracy. However, technical barriers like cumbersome implementation in established workflows hinder their application in practice. With MLPROP, we provide an interactive web interface for directly applying advanced ML methods to predict thermophysical properties without requiring ML expertise, thereby substantially increasing the accessibility of novel models. MLPROP currently includes models for predicting the vapor pressure of pure components (GRAPPA), activity coefficients and vapor-liquid equilibria in binary mixtures (UNIFAC 2.0, mod. UNIFAC 2.0, and HANNA), and a routine to fit NRTL parameters to the model predictions. MLPROP will be continuously updated and extended and is accessible free of charge via \url{https://ml-prop.mv.rptu.de/}. MLPROP removes the barrier to learning and experimenting with new ML-based methods for predicting thermophysical properties. The source code of all models is available as open source, which allows integration into existing workflows.
\end{abstract}

\section{Introduction}
Knowledge of the thermophysical properties of fluids is essential in many fields, e.g., for designing and optimizing chemical processes. Since experimental data are often missing, reliable prediction methods are indispensable. Therefore, many physical and semi-empirical methods for predicting various thermophysical properties of pure fluids and mixtures have been developed \cite{Poling2001}, some of which are widely used in industrial practice \cite{Gmehling2019}. However, these methods are often limited in their scope and the accuracy of their predictions. Recently, machine learning (ML) has opened the route to develop new methods for predicting thermophysical properties that often outperform the established methods in application range and prediction accuracy \cite{Jirasek2021, Jirasek2023, Hoffmann2025, Specht2024, Hayer2025, Hayer2024, SanchezMedina2022, Winter2022, Winter2025}. While, in many cases, the codes of the models are freely available, hurdles must be overcome before they can be executed or even implemented in industrial workflows, e.g., installing several software packages that may not immediately run on the available hardware. Quick and easy access to the new models is essential to deciding whether it is worthwhile to overcome these barriers. 

In order to create trust in the new ML-based methods for the prediction of thermophysical properties and to accelerate their acceptance and implementation in the workflows in industry and academia, we introduce MLPROP. This intuitive web interface makes some of the most advanced ML methods in this field accessible without requiring expertise in ML or coding. Currently, MLPROP integrates several hybrid ML models that allow the prediction of vapor pressures, boiling temperatures, activity coefficients, and vapor-liquid equilibria. This paper briefly summarizes the implemented models and gives an introduction to the use of MLPROP.

\section{Implemented models}
This section gives a short overview of the models currently implemented in MLPROP: GRAPPA \cite{Hoffmann2025} for predicting component-specific Antoine parameters from which vapor pressures and boiling temperatures can be calculated; UNIFAC 2.0 \cite{Hayer2025}, mod. UNIFAC 2.0 \cite{Hayer2024}, and HANNA \cite{Specht2024} for predicting activity coefficients in binary systems. We refer the reader to the original publications for more detailed information on the models, their architecture, training data, and comprehensive benchmarking with methods from the literature. 

\subsection{GRAPPA} 
GRAPPA\cite{Hoffmann2025} is a hybrid graph neural network for predicting vapor pressures of pure organic components. GRAPPA predicts the parameters of the Antoine equation (\textit{A}, \textit{B}, \textit{C})  from the molecular structure, which can subsequently be used for predicting the vapor pressure as a function of temperature or vice versa, the boiling temperature as a function of pressure. GRAPPA was trained on more than 200,000 experimental data points of more than 20,000 pure components. As GRAPPA needs SMILES as the only input, it has broad applicability. Its predictive power was thoroughly tested using components not used for the training and was found to outperform several methods from the literature~\cite{Tu1994, Pankow2008, Nannoolal2008, Lin2024}.

\subsection{UNIFAC 2.0 and mod. UNIFAC 2.0}
UNIFAC 2.0 \cite{Hayer2025} and mod. UNIFAC 2.0 \cite{Hayer2024} are enhanced versions of the widely established group contribution methods UNIFAC \cite{Fredenslund1975, Wittig2002} and mod. UNIFAC (Dortmund) \cite{Weidlich1987, Constantinescu2016}, respectively. A key element of all UNIFAC versions is the parameter table with the numerical values of the group-interaction parameters. While the original versions have significant gaps in their parameter tables, these are filled in the enhanced 2.0 versions, significantly extending the applicability of UNIFAC. However, not only is the scope widened by the enhanced versions, but the quality of the predictions has also improved. Developing UNIFAC 2.0 and mod. UNIFAC 2.0 was achieved by embedding matrix completion methods \cite{Jirasek2020, Jirasek2022, Jirasek2023a} into the framework of the physical models and training them to predict pair-interaction parameters for which no experimental data for direct fitting are available. By completely retaining the physical framework of the original models, full thermodynamic consistency is contained in the hybrid models. UNIFAC~2.0 and mod. UNIFAC~2.0 were trained on more than 200,000 data points for activity coefficients in binary mixtures and, in the case of mod. UNIFAC 2.0, additionally on over 250,000 data points for excess enthalpies of binary mixtures.

\subsection{HANNA} 
HANNA \cite{Specht2024} is the first neural network for predicting activity coefficients in which all relevant consistency criteria are hard-coded so that the predictions strictly comply with them. In its present version, HANNA is restricted to binary systems. However, an extension to multi-component systems is being developed and will be included in MLPROP shortly. HANNA requires only the SMILES of the two components as well as the temperature and information on the composition of the mixture as input and yields the two activity coefficients. HANNA was trained on more than 300,000 experimental data points of activity coefficients in binary mixtures and achieves significantly higher accuracy than mod. UNIFAC (Dortmund) \cite{Weidlich1987, Constantinescu2016} on a test set of unseen mixtures. In MLPROP, we provide two versions of HANNA: the original version reported in a recent paper \cite{Specht2024} and a prototype ensemble model with slightly better accuracy.

\section{Using MLPROP}
On the MLPROP homepage \url{https://ml-prop.mv.rptu.de}, the user is guided in a dialog in which the first step is selecting a prediction task. Then, the available models for that task are listed for the user to choose from. Finally, the user needs to enter the molecular structure of the component(s) of interest in the SMILES notation and, depending on the prediction task, the temperature or pressure. If the user does not have the SMILES notation of the component(s) at hand, we recommend the use of PubChem, a public chemical database resource at the National Institute of Health, for the conversion of IUPAC names, CAS registry numbers or other chemical identifiers into SMILES notation. The service is available under \url{https://pubchem.ncbi.nlm.nih.gov/}. MLPROP automatically checks the entered SMILES for validity and canonizes them.

\subsection{Prediction of vapor pressures and boiling points of pure components}
The vapor pressure and boiling temperature predictions for pure components are based on GRAPPA\cite{Hoffmann2025}. MLPROP can provide either the parameters of the Antoine equation or the vapor pressure for a given temperature or the boiling temperature for a given pressure. The considered components must fall into the chemical space covered by GRAPPA, e.g., the components should have at least two heavy atoms and one carbon atom. Furthermore, it should be considered that the uncertainty of the predictions increases for pressures below 1 kPa, where often only poor experimental data are available. For more details, see Ref.~\cite{Hoffmann2025}. Note that the present version of GRAPPA has no information on the critical point and the triple point, so it will also yield predictions outside the limits of the actual vapor-pressure curve.

\subsection{Prediction of activity coefficients in binary mixtures}
The user can choose between UNIFAC 2.0, mod. UNIFAC 2.0, HANNA, and the prototype ensemble version of HANNA for predicting activity coefficients in the liquid phase of binary mixtures. In all cases, activity coefficients of both components over the complete composition range are calculated in $0.01\,\mathrm{mol}\,\mathrm{mol}^{-1}$ increments for a constant temperature, which the user can specify. (Note that the routine does not check whether all states at that temperature are liquid). The results are displayed in an interactive plot, which can be downloaded as \textit{png} file or in the \textit{csv} format. For the UNIFAC-based methods, the chemical space is limited to components that can be decomposed into the defined structural groups according to the Dortmund Data Bank \cite{DDB2024}. In theory, all valid SMILES are also valid inputs for HANNA. However, predictions for unusual components or extreme conditions should be interpreted cautiously. 

\subsection{Prediction of NRTL parameters}
MLPROP provides a routine to fit the NRTL model \cite{Renon1968} to the numerical data obtained from the methods described in Section 3.2 to simplify the integration of the predicted activity coefficients into established process simulators. The fit is performed using the \verb|minimize()| function of the \textit{scipy} python package to minimize the fit error
\begin{equation}
    \mathcal{L} = \frac{1}{2NJ}\sum_{i=1}^N \sum_{j=1}^J\sum_{k=1}^2 \left(\ln \gamma_k^\mathrm{NRTL}(x_i, T_j) - \ln \gamma_k^\mathrm{pred}(x_i, T_j) \right)^2
\end{equation}
Here, $N$ is the number of discrete compositions, $J$ is the number of different temperatures, and $k$ marks components 1 or 2. The superscript 'pred' denotes the predicted activity coefficients from the employed ML model. In the current version of MLPROP, fits of NRTL model variants with three, six, and ten parameters are available. These model variants were chosen based on the NRTL implementation in Aspen Plus V12 \cite{AspenTech2020}]. The exact equations are displayed if the user fits the NRTL model. For the three-parameter fit, the composition space is discretized with $\Delta x_i=0.01\,\mathrm{mol}\,\mathrm{mol}^{-1}$ ($N=101$) and only isothermal fits are possible since this approach is the least flexible. For the six- and ten-parameter options, a temperature range can be specified, which is discretized equally spaced with $J=5$, and the discretization in the composition space is set to $N=21$. The resulting NRTL parameters and underlying equations corresponding with the fit are displayed below the plot. The fitted parameters and the NRTL predictions can also be downloaded as \textit{csv} file. We note that it is at the user's discretion to assess the fit's quality and choose the appropriate parameter form for the system and temperature range of interest.

\subsection{Prediction of vapor-liquid equilibria}
The prediction of vapor-liquid equilibria (VLE) of binary mixtures is based on the extended Raoult's law
\begin{equation}
    p_i^s(T)\,x_i\,\gamma_i(T,\bm{x}) = p\,y_i 
    \label{eq:raoults_law}
\end{equation}
where $p_i^\mathrm{s}(T)$ is the vapor pressure of component \textit{i} at the temperature \textit{T}, $x_i$ and $y_i$ are the mole fractions of component \textit{i} in the liquid and vapor phase in equilibrium, respectively, $\gamma_i(T,\bm{x})$ is the activity coefficient of component \textit{i} in the liquid phase of the binary mixture, and $p$ is the pressure. The vapor pressures are calculated with GRAPPA. For the activity coefficients, the user can choose between the available methods (UNIFAC 2.0, mod. UNIFAC~2.0, ensemble HANNA). 

The extended Raoult's law assumes an ideal gas phase and neglects the Poynting correction. These assumptions are reasonable up to moderate pressures. Based on Eq.~(\ref{eq:raoults_law}), MLPROP predicts the entire vapor-liquid phase diagram either for constant temperature $T$ or constant pressure $p$, according to the specification by the user. Consequently, MLPROP calculates the bubble and dew line (i.e., the corresponding $p$ in the isothermal case and the corresponding $T$ in the isobaric case) over the entire composition space discretized in increments of 0.01 mol/mol, i.e., the liquid phase composition (bubble line) or vapor phase composition (dew line) is specified. In the isothermal case, the user obtains the $p$-$x$-$y$ diagram and a diagram of the corresponding logarithmic activity coefficients. In the isobaric case, the $T$-$x$-$y$ diagram and the corresponding $x$-$y$ diagram are displayed. 

The current implementation does not support the calculation of VLE with immiscible liquid phases, i.e., the prediction of heteroazeotropes is not feasible. To avoid the display of incorrect results, the constructed VLE diagrams are automatically checked for thermodynamic consistency, i.e., that bubble and dew line merge in the pure components, the derivatives of bubble and dew line have the same sign, the bubble line always lies above the dew line (isothermal case) or vice versa (isobaric case), and bubble and dew line coincide in an azeotropic point and their derivatives have a sign change there\cite{Gmehling2019}. If any of these criteria is not fulfilled, MLPROP will issue an error. Note that because GRAPPA does not predict the triple point and critical point of the pure components, the predicted VLE diagrams for state points that do not lie within the actual vapor pressure curves of the components can look reliable but are not physically meaningful. 

\begin{figure}[h!]
    \centering
    \includegraphics[width=0.91\linewidth]{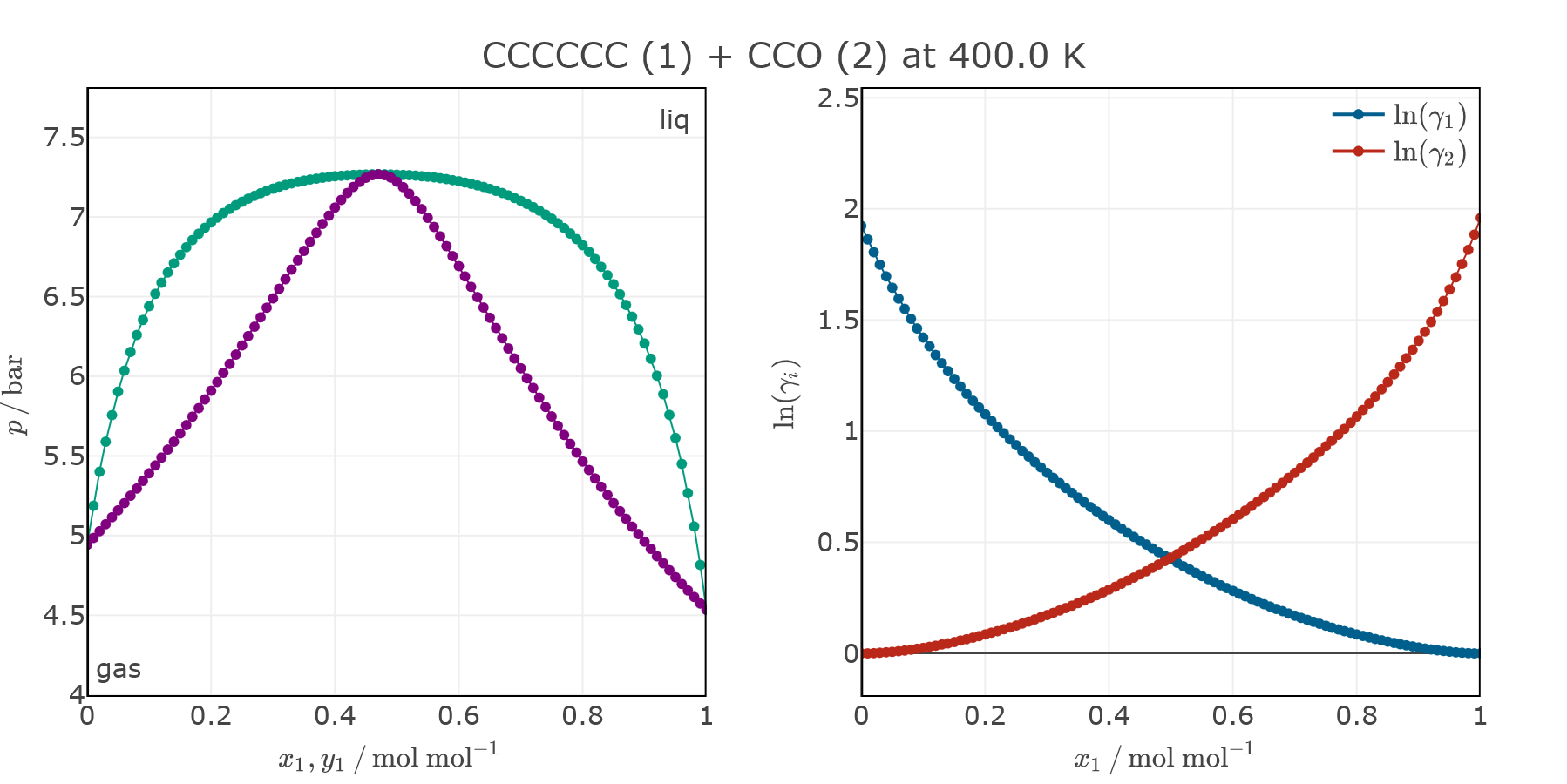}
    \caption{Downloaded MLPROP screenshot of the isothermal VLE of hexane~(1) and ethanol~(2) at $T=400\,\mathrm{K}$ predicted with GRAPPA and ensemble HANNA. Left: $p$-$x$-$y$ phase diagram including the bubble line (green) and dew line (purple). Right: Logarithmic activity coefficients in the binary mixture obtained from ensemble HANNA over the mole fraction of hexane.}
    \label{fig:vle_isotherm}
\end{figure}

\begin{figure}[h!]
    \centering
    \includegraphics[width=0.91\linewidth]{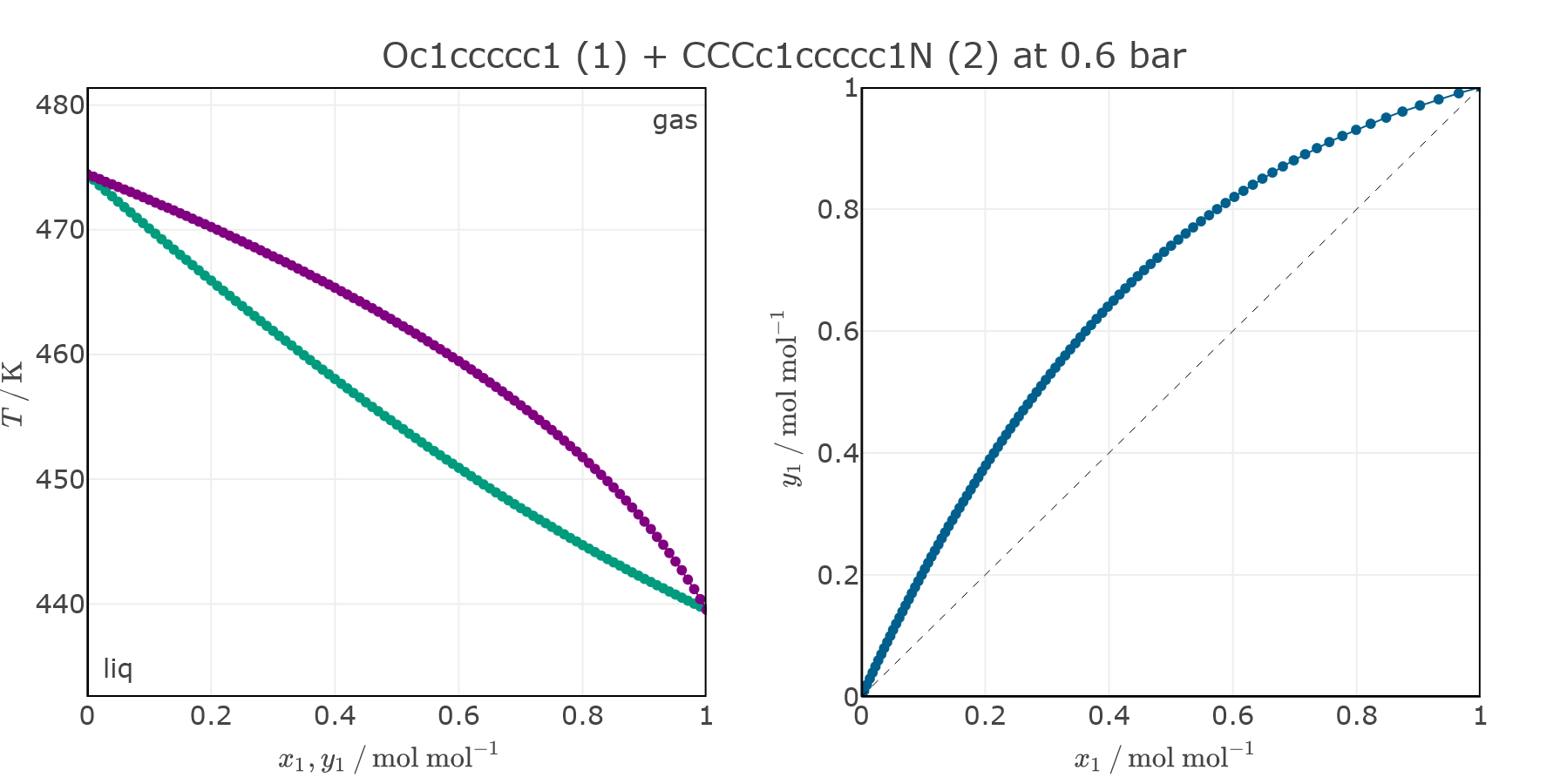}
    \caption{Downloaded MLPROP screenshot of the isobaric VLE of phenol~(1) and 2-propylaniline~(2) at $p=0.6\,\mathrm{bar}$ predicted with GRAPPA and mod.~UNIFAC 2.0. Left: $T$-$x$-$y$ phase diagram including the bubble line (green) and dew line (purple). Right:~Corresponding $x$-$y$ diagram.}
    \label{fig:vle_isobaric}
\end{figure}

The phase diagrams supplied by MLPROP can be downloaded as \textit{png} and/or \textit{csv} file. In Fig.~\ref{fig:vle_isotherm}, the output of MLPROP for the isothermal VLE of hexane (CCCCCC) and ethanol (CCO) at $T=400\,\mathrm{K}$ is displayed, Fig.~\ref{fig:vle_isobaric} shows the output for the isobaric VLE of phenol (Oc1ccccc1) and 2-propylaniline (CCCc1ccccc1N) at $p=0.6\,\mathrm{bar}$.

\section{Conclusions and outlook}
We introduce MLPROP, an interactive web interface that enables easy access to advanced ML methods for predicting the thermophysical properties of pure components and mixtures. MLPROP can immediately be used without any detailed knowledge of the ML methods, and their performance can be tested for any problem within their scope. All implemented methods can easily be integrated into fully automated workflows, as their source code is publicly available. Currently, MLPROP includes hybrid ML models for vapor pressures and activity coefficients that enable the purely predictive calculation of vapor-liquid equilibria. It will be continuously updated and extended as new models become available. MLPROP is maintained and hosted at the Laboratory of Engineering Thermodynamics (LTD) at RPTU Kaiserslautern. The developers appreciate user feedback to guide future developments of ML methods and their integration into MLPROP. 

\section*{Acknowledgments}
We gratefully acknowledge financial support by the Carl Zeiss Foundation in the project "Process Engineering 4.0" and by DFG in the Priority Program SPP2363 "Molecular Machine Learning" (grant number 497201843). Furthermore, FJ gratefully acknowledges financial support by DFG in the frame of the Emmy-Noether program (grant number 528649696).

\bibliographystyle{unsrt}  
\bibliography{references}

\end{document}